\documentclass[10pt,conference]{IEEEtran}

\makeatletter
\def\ps@headings{%
\def\@oddhead{\mbox{}\scriptsize\rightmark \hfil \thepage}%
\def\@evenhead{\scriptsize\thepage \hfil \leftmark\mbox{}}%
\def\@oddfoot{}%
\def\@evenfoot{}}
\makeatother

\usepackage{amsfonts}
\usepackage{mathrsfs}
\usepackage{amsfonts}
\usepackage{amssymb}
\usepackage{graphicx}
\usepackage{epsfig}
\usepackage{epstopdf}
\usepackage{psfrag}
\usepackage{amsmath}
\usepackage{array}
\usepackage{subfigure}
\usepackage{cite,graphicx,amsmath,amssymb,color}
\usepackage{anysize}
\usepackage{algorithmic}
\usepackage{algorithm}

\usepackage{algorithm}
\usepackage{algorithmic}
\usepackage{stmaryrd}
\usepackage{multirow}
\usepackage{subfig}
\usepackage{graphicx,times,amsmath} 

\usepackage{url}
\usepackage{enumerate}

\marginsize{15.8mm}{15.8mm}{19.1mm}{25.4mm}

\IEEEoverridecommandlockouts

\newtheorem{theorem}{Theorem}
\newtheorem{lemma}{Lemma}

\newtheorem{Def}{Definition}

\newtheorem{problem}{Problem}
\newtheorem{remark}{Remark}

\begin{document}
\bibliographystyle{IEEEtran}

\title{Energy-Efficient Resource Allocation for Cache-Assisted  Mobile Edge Computing}
\author{\authorblockN{Ying Cui\thanks{The work of Y. Cui was supported by NSFC grant 61401272 and grant 61521062.
The work of Z. Liu was supported by JSPS KAKENHI Grant Numbers 16H02817 and 15K21599.}, Wen He, Chun Ni, Chengjun Guo} \authorblockA{Department of Electronic Engineering\\Shanghai Jiao Tong University, China}\and\authorblockN{Zhi Liu}\authorblockA{Department of Mathematical and Systems Engineering\\Shizuoka University, Japan}}
\maketitle

\begin{abstract}
In this paper, we jointly consider  communication, caching and computation  in a multi-user cache-assisted mobile edge computing (MEC)  system,   consisting of  one base station (BS) of caching and computing capabilities and multiple users with computation-intensive and latency-sensitive  applications. 
We propose a joint caching  and  offloading mechanism which involves task  uploading and executing for tasks with uncached computation results as well as computation result downloading for all tasks at the BS, and efficiently utilizes multi-user diversity   and multicasting opportunities.
Then, we formulate the average total energy   minimization problem subject to the caching and deadline constraints  to optimally allocate the storage resource at the BS  for caching computation results as well as the  uploading and downloading time durations. The problem is a challenging mixed discrete-continuous optimization problem. We  show that strong duality holds, and obtain an optimal solution using a dual method. To reduce the computational complexity, we further propose a low-complexity suboptimal solution.
Finally, numerical results show that the proposed suboptimal solution outperforms existing comparison schemes.
\end{abstract}


\section{Introduction}\label{sec:intro}

With drastic development of mobile devices, new applications with advanced features such as augmented reality, mobile online gaming and multimedia
transformation, are emerging. These listed applications are both latency-sensitive and computation-intensive, and are beyond the computing capability of common mobile devices.  Mobile edge computing (MEC) is one promising technology which provides the computing capability to support these applications at the wireless edge.
In an MEC system, a mobile user's computation task can be uploaded to a base station (BS) and executed at its attached MEC server, which significantly releases the mobile user's computation burden.
However, at the wireless edge, limited communication and computation resources bring big challenges for MEC systems to satisfy massive demands for these applications~\cite{mao2017mobile}.
Designing energy-efficient MEC systems requires  a joint optimization of communication and computation resources   among distributed mobile devices and MEC servers. Such optimal resource allocation  has been considered  for various types of multi-task MEC systems~\cite{mao2016dynamic,you2016energy,wang2017joint,Guo2017,chen2016joint}. For instance,   \cite{mao2016dynamic,you2016energy,wang2017joint,Guo2017} study a multi-user MEC system with one BS and one inelastic task for each user, and minimize the energy consumption under {a hard deadline constraint for each task}. In \cite{chen2016joint}, the authors investigate a multi-user MEC system with one BS and multiple independent elastic tasks for each user, and consider the minimization of the overall system cost. In particular, the offloading scheduling \cite{mao2016dynamic,wang2017joint,chen2016joint} and transmission time (or power) allocation \cite{you2016energy,wang2017joint,Guo2017,chen2016joint} are considered in these optimizations.

One common assumption adopted in~\cite{mao2016dynamic,you2016energy,wang2017joint,chen2016joint,Guo2017} is that the computation tasks for different mobiles are different and the computation results cannot be reused, which may not always hold in practice.
For instance, in augmented reality subscriptions for better viewing experience in museums, a processed augmented reality output may be simultaneously or asynchronously   used by visitors in the same place \cite{mao2017mobile}.
Another example is  mobile online game where a processed gaming scene may be  requested synchronously by a group of   players  or asynchronously by individual players. In these scenarios where task requests are highly concentrated in the spatial domain and asynchronously or synchronously  repeated in the time domain~\cite{al2016optimal,tran2017collaborative},  storing  computation results closer to users (e.g., at BSs) for future reuse can greatly reduce the  computation burden and   latency.
For example, in \cite{al2016optimal}, the authors  propose a resource allocation approach which allows   users to share computation results, and minimize the total mobile energy consumption  for offloading under the  latency and power constraints. However, this paper focuses on only one computation task  and does not consider caching computation results for future demands. The authors of \cite{tran2017collaborative} propose collaborative multi-bitrate video caching and processing in a  multi-user MEC system to minimize the backhaul  load, without considering the energy consumption for  task executing and computation result downloading.  To the best of our knowledge, how to design energy-efficient cache-assisted MEC systems by  jointly  optimizing communication, caching and computation resources remains unsolved.

In this paper, we jointly consider  communication, caching and computation in a multi-user cache-assisted MEC system consisting of  one BS of caching and computing capabilities and multiple users with inelastic computation tasks.
We specify each task using three parameters, i.e., the size of the task input, workload and size of the computation result. In addition, we consider  the  popularity  and  the randomness in  task requirements.
Based on this task model, we propose a caching and  offloading mechanism  which involves task  uploading and executing for tasks with uncached computation results as well as computation result downloading for all tasks at the BS, and efficiently utilizes  multi-user diversity in task uploading and multicasting opportunities in computation result  downloading. Then, we formulate the average total energy   minimization problem  subject to the caching and  deadline constraints  to optimally allocate the storage resource at the BS  as well as the uploading and downloading time durations. The problem is a challenging mixed discrete-continuous optimization problem. We convert its dual problem to a knapsack problem for caching and multiple  convex problems for uploading and downloading time allocation, and obtain the dual optimal solution using  the subgradient method. We also show that strong duality holds,  and obtain an optimal solution of the primal problem based on the dual optimal solution. To reduce the computational complexity, we further propose a low-complexity suboptimal solution.
Finally, numerical results show that the proposed suboptimal solution  outperforms existing comparison schemes.

\begin{figure}[t]
\begin{center}
  \subfigure
  {\resizebox{8cm}{!}{\includegraphics{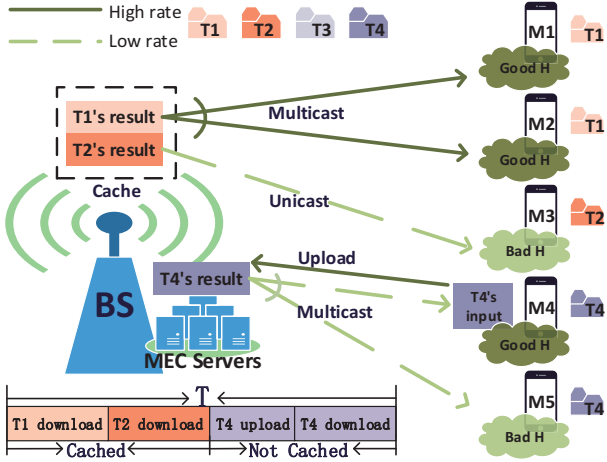}}}
  \end{center}
  \caption{\small{ System model. Note that  T, M and H are short for task, mobile and channel, respectively. $K=5$, $N=4$ and $|\mathcal H|=2$.}}
\label{system_model}
\end{figure}

\section{System Model}


As illustrated in Fig.~\ref{system_model}, we consider a multi-user  cache-assisted MEC system with one BS and $K$ single-antenna mobiles, denoted by
set $\mathcal K \triangleq \{1, 2, ...,K\}$. The MEC system operates on a frequency band with a bandwidth $B$ (Hz). The BS has powerful caching and computing capabilities
at the network edge.
Each mobile has a  computation-intensive and latency-sensitive computation task which is generated at time 0 and has deadline $T$ (in seconds), and is offloaded to the BS for executing (due to crucial
computation and latency requirements). We consider the operation of  the MEC system in  time interval $[0,T]$.
Note
that for each user, multiple tasks which are generated at the same time and have the same deadline can be viewed
as one super-task whose  workload is the sum of the workloads of all its  task components. We would like to obtain first-order design insights into caching and computing in cache-assisted MEC. The results obtained in
this paper can be extended to study a more general scenario where
some tasks can be executed locally and different tasks may have different deadlines.
\subsection{Task Model and Channel Model}
Consider $N$ computation-intensive and latency-sensitive computation tasks, denoted by set $\mathcal X\triangleq \{1,\cdots, N\}$. As in \cite{Guo2017},
each task $n\in\mathcal X$ is characterized by three parameters, i.e., the size of the task input $L_{u,n}>0$ (in bits), workload $L_{e,n}>0$ (in number of CPU-cycles), and size of the computation result  $L_{d,n}>0$ (in bits).  The computation result of each task has to be obtained  within $T$ seconds.
Note that the three parameters of a
computation task are determined by the nature of the task
itself, and   can be estimated
to certain extent based on some prior offline measurements \cite{Guo2017}. In addition, the adopted task model  properly addresses the limitation in  prior work  that computation results are assumed to be negligible in size and  trivial to download\cite{mao2016dynamic,you2016energy}.

Different from \cite{Guo2017}, we focus on the scenario where one task may be required  by multiple users, and hence its computation result can be reusable.  Examples of these types of applications have been illustrated in Section~\ref{sec:intro}.  To reflect this characteristic, we model the task popularity.
Specifically, mobile $k$ needs to execute a random computation task, denoted by $X_k\in\mathcal X$. Let $p_{X_k}(x_k)\triangleq \Pr[X_k=x_k]\geq 0$ denote the probability that the random variable $X_k$ takes the value $x_k\in \mathcal X$. Note that $\sum\limits_{x_k\in \mathcal X}p_{X_k}(x_k)=1$.   Suppose the discrete random variables  $X_k,k\in\mathcal K$ are independently  distributed, and their probability mass functions (p.m.f.s) $p_{X_k}(\cdot)$, $k\in\mathcal K$ can be different.
Let $\mathbf X\triangleq (X_k)_{k\in\mathcal K}\in \mathcal X^K$ denote the random system task state.

We consider a block fading model for wireless channels. Let $H_k\in \mathcal H$ denote the random channel state of mobile $k$, representing the power of the channel  between mobile $k$ and the BS,  where $\mathcal H$ denotes the finite channel  state space. Assume  $H_k$ is constant during the $T$ seconds. Let
$p_{H_k}(h_k)\triangleq \Pr[H_k=h_k]\geq 0$ denote the probability that the random variable $H_k$ takes the value $h_k\in\mathcal H$. Note that  $\sum\limits_{h_k \in \mathcal H}p_{H_k}(h_k)=1$.
Suppose  the discrete random variables $H_k,k\in\mathcal K$ are independently distributed, and their p.m.f.s $p_{H_k}(\cdot)$, $k\in\mathcal K$ can be different.
Let $\mathbf H\triangleq (H_k)_{k\in\mathcal K}\in \mathcal H^{K}$ denote the random system channel state.

The random system state consists of  the random system task state $\mathbf X$ and the random system channel state $\mathbf H$, denoted by $(\mathbf X, \mathbf H)\in  \mathcal X^K\times  \mathcal H^K$. Suppose $\mathbf X$ and  $\mathbf H$ are independent. Thus, the probability that the random system state $(\mathbf X,\mathbf H)$ takes the value $(\mathbf x,\mathbf h)\in  \mathcal X^K\times  \mathcal H^K$ is given by
\begin{align}
\Pr[(\mathbf X,\mathbf H)=(\mathbf x, \mathbf h) ]=\prod_{k\in\mathcal K}p_{X_k}(x_k)p_{H_k}(h_k)\triangleq p(\mathbf x,\mathbf h),
\end{align}
where $\mathbf x\triangleq (x_k)_{k\in\mathcal K}\in \mathcal X^K$ and $\mathbf h \triangleq (h_k)_{k\in\mathcal K}\in \mathcal H^K$.
Each mobile can inform the BS the  I.D.  of the task  it needs to execute, and the BS can easily obtain the channel state of each mobile (e.g., by channel sounding). Thus,  we assume that the BS is aware of the system state $(\mathbf X,\mathbf H)$.

Let $\mathcal K_n(\mathbf X)\triangleq \{k\in\mathcal K: X_k=n\}$ and $K_n(\mathbf X)\triangleq \sum_{k \in \mathcal K}\text{I}\left[X_k=n\right]$ denote the set and number of mobiles who need to execute task $n$ at the random system task state $\mathbf X$, where $\text{I}[\cdot]$ denotes the indicator function.   Note that $K_n(\mathbf X)=|\mathcal K_n(\mathbf X)|$.  When  there exists at least one user requiring to execute  task $n$, i.e., $K_n(\mathbf X)\geq 1$,  let $H_{u,n}$ and $H_{d,n}$ denote the largest and smallest values among the channel states of all the $K_n(\mathbf X)$  mobiles in $\mathcal K_n(\mathbf X)$, respectively, where
\begin{align}
H_{u,n}\triangleq &\max_{k\in\mathcal K_n(\mathbf X)} H_k,\ n\in\mathcal X,\label{eqn:best-H}\\
H_{d,n}\triangleq &\min_{k\in\mathcal K_n(\mathbf X)} H_k, \ n\in\mathcal X.\label{eqn:worst-H}
\end{align}
Note that $H_{u,n}$ and $H_{d,n}$ are  determined by  $(\mathbf X,\mathbf H)$.

\subsection{Caching and Offloading}
First, we consider caching reusable computation results.
The BS is equipped with a cache of size $C$ (in bits), and can store some computation results.
Let $c_n$  denote the caching action for  the computation result of task $n$ at the BS, where
\begin{align}
&c_n \in \{0,1\}, \; n \in \mathcal X. \label{caching}
\end{align}
Here,  $c_n=1$ means that the computation result of task $n$ is cached, and $c_n=0$ otherwise.
Under the cache size constraint at the BS, we have
\begin{align}
&\sum_{n\in \mathcal X}c_nL_{d,n}\leq C.\label{eqn:cache-const}
\end{align}

Next, we introduce task offloading. The BS is of computing capability by running  a server of a constant CPU-cycle frequency  and can execute computation tasks from mobiles. Consider two scenarios in offloading task $n$ to the BS for executing, depending on whether the computation result of task  $n$ is stored at the BS or not.
If the computation result of task $n$ is not cached at the BS, i.e.,  $c_n=0$,  offloading task $n$   to the BS for executing comprises three sequential stages: 1) uploading the input of task $n$ with $L_{u,n}$ bits from the mobile with the best channel $H_{u,n}$ among all the $K_n(\mathbf X)$ mobiles in $\mathcal K_n(\mathbf X)$ to the BS; 2) executing task $n$  at the BS (which requires $L_{e,n}$ CPU-cycles); 3) downloading  the computation result with $L_{d,n}$ bits from the BS to all the $K_n(\mathbf X)$ mobiles in $\mathcal K_n(\mathbf X)$ using multicasting.   Note that both the uploading and downloading are over the whole frequency band. Recall that the BS is aware of the system state $(\mathbf X,\mathbf H)$.
In uploading  the input of task $n$, instead of letting each of the $K_n(\mathbf X)$ mobiles in $\mathcal K_n(\mathbf X)$ upload separately, the BS selects the mobile with the best channel $H_{u,n}$ to upload. This wisely avoids redundant transmissions and fully makes use of multi-user diversity, leading to  energy reduction in uploading. In addition,  in downloading the computation result of task $n$, the BS transmits only once at a certain rate so that the mobile with the worst channel $H_{d,n}$  can successfully receive the computation result.  Let $t_{u,n}$ denote the downloading time duration   for task $n$, where
\begin{align}
0 \leq t_{u,n}\leq T, \ n \in \mathcal X.\label{eqn:time_constraint_1_}
\end{align}
The BS executing time (in seconds) for task $n$ is  $t_{e,n} = L_{e,n}/F_b$,
where  $F_b> 0$ denotes the fixed CPU-cycle frequency of the BS.  As $F_b$ is usually large, $t_{e,n}$ is small. In the following, for ease of analysis, we ignore the BS  executing time, i.e., assume $t_{e,n}=0$ ~\cite{wang2017joint}.
Let $t_{d,n}$ denote the downloading time duration for task $n$, where
\begin{align}
0 \leq t_{d,n}\leq T,\ n \in \mathcal X.\label{eqn:time_constraint_2_}
\end{align}
If the computation result of task $n$ is cached at the BS, i.e., $c_n=1$,  directly  offloading task $n$ to the BS for executing involves only  one stage, i.e., downloading  the computation result of task $n$ from the BS  to all the $K_n(\mathbf X)$ mobiles in $\mathcal K_n(\mathbf X)$  using multicasting, with the downloading time duration satisfying \eqref{eqn:time_constraint_2_}.

We consider Time Division Multiple Access (TDMA) with Time-Division Duplexing (TDD) operation \cite{Guo2017,you2016energy,mao2016dynamic,wang2017joint}. Note that when the BS executing time is negligible, the processing order for the offloaded tasks does not matter \cite{Guo2017}, and the total completion time is the sum of the uploading time durations of the tasks whose computation results are not cached and the downloading time durations of the computation results of  all  tasks. Thus,  under the deadline constraint, we have
\begin{align}
\sum_{n\in\mathcal X}\left((1-c_n)t_{u,n} + t_{d,n} \right)\leq T. \label{eqn:time_constraint_3_}
\end{align}

\subsection{Energy Consumption}
We now introduce the transmission energy consumption model for  uploading and downloading.    First, consider $c_n=0$. Recall that in this case, the mobile with the best channel among all the $K_n(\mathbf X)$ mobiles in $\mathcal K_n(\mathbf X)$ uploads task $n$ to the BS. Let $p_{u,n}$ denote the transmission power. Then, the achievable transmission rate (in bit/s)  is  $$r_{u,n} = B \log_2 \left(1 + \frac{p_{u,n} H_{u,n}}{n_0}\right),$$
where $B$ and  $n_0$ are the bandwidth and the power of the complex additive  white Gaussian noise, respectively. On the other hand, the transmission rate should be fixed as $r_{u,n} = L_{u,n}/t_{u,n}$, since this is the most energy-efficient transmission method for transmitting $L_{u,n}$ bits in $t_{u,n}$ seconds (due to the fact that $$p_{u,n}=\frac{n_0}{H_{u,n}}(2^{r_{u,n}/B}-1)$$ is a convex function of $r_{u,n}$).
Define $$g(x) \triangleq n_0 \left(2^{\frac{x}{B}} - 1\right).$$ Then, we have $p_{u,n} = \frac{1}{H_{u,n}}g\left(\frac{L_{u,n}}{t_{u,n}}\right)$. Thus, at the system state $(\mathbf X, \mathbf H) $, the transmission energy consumption for uploading the input of  task $n$ to the BS with the uploading time duration $t_{u,n}$  is given by:
\begin{align}\label{eqn:energy-u}
& E_{u,n}(t_{u,n},\mathbf X, \mathbf H)
\triangleq \begin{cases}
\frac{t_{u,n}}{H_{u,n}} g\left(\frac{L_{u,n}}{t_{u,n}}\right), & K_n(\mathbf X)\geq 1 \\
0, &\text{otherwise}
\end{cases} \ n\in\mathcal X,
\end{align}
where $H_{u,n}$ is given by \eqref{eqn:best-H}.
In addition, recall that the BS multicasts the computation result of task $n$ to all the $K_n(\mathbf X)$ mobiles in $\mathcal K_n(\mathbf X)$.
Thus, similarly, at the system state $(\mathbf X, \mathbf H) $,  the transmission energy consumption at the BS for multicasting  the computation result of task $n$  with the downloading time duration $t_{d,n}$  is given by:
\begin{align}
&E_{d,n}(t_{d,n},\mathbf X,\mathbf H) \triangleq \begin{cases}
\frac{t_{d,n}}{H_{d,n}}g\left(\frac{L_{d,n}}{t_{d,n}}\right), &  K_n(\mathbf X)\geq 1\\
0, &\text{otherwise}
\end{cases}\ n\in\mathcal X,\label{eqn:energy-d}
\end{align}
where $H_{d,n}$ is given by \eqref{eqn:worst-H}.
Then, consider $c_n=1$. In this case,   the BS directly multicasts the computation result of task $n$ stored at the BS to all the $K_n(\mathbf X)$ mobiles in $\mathcal K_n(\mathbf X)$ with the  transmission energy $E_{d,n}(t_{d,n},\mathbf X,\mathbf H) $ given in \eqref{eqn:energy-d}.

Next, we illustrate  the computation energy consumption at the BS.  
We consider low CPU voltage of the server at the BS.  The  energy consumption for computation in a single CPU-cycle with frequency $F_b$ is $\mu F_b^2$, where $\mu$ is a constant factor determined by the switched capacitance of the server~\cite{you2016energy}. Then, the energy consumption for executing task $n$ at the BS is:
\begin{align}
    &E_{e,n} (\mathbf X)\triangleq
    \begin{cases}
    \mu L_{e,n} F_b^2, & K_n(\mathbf X)\geq 1\\
    0, &\text{otherwise}
\end{cases}\ n\in\mathcal X.\label{eqn:energy-b}
\end{align}

Therefore, the  energy consumption for task $n$ is given by\footnote{Note that by multiplying $E_{e,n}(\mathbf X)$ and $E_{d,n}(t_{d,n},\mathbf X,\mathbf H)$ with a scalar in interval $(0,1)$, different weights for the energy consumptions at the BS and the mobiles can be reflected. The proposed framework can be easily extended.}
\begin{align}
&E_n(c_n, t_{u,n},t_{d,n}, \mathbf X,\mathbf H)\nonumber\\
\triangleq&    (1-c_n)(E_{u,n}(t_{u,n},\mathbf X,\mathbf H)+E_{e,n}(\mathbf X))+E_{d,n}(t_{d,n},\mathbf X,\mathbf H).\label{eqn:energy-X-H-n}
\end{align}
Then, the total energy consumption is  given by
\begin{align}
&E(\mathbf c, \mathbf t_u,\mathbf t_d, \mathbf X,\mathbf H)\triangleq  \sum_{n\in \mathcal X}E_n(c_n, t_{u,n},t_{d,n}, \mathbf X,\mathbf H), \label{eqn:energy-X-H}
\end{align}
where $\mathbf c \triangleq (c_n)_{n \in \mathcal X}$, $\mathbf t_u \triangleq (t_{u,n})_{n \in \mathcal X}$ and $\mathbf t_d \triangleq (t_{d,n})_{n \in \mathcal X}$. 

\section{Problem Formulation}

Define the feasible  joint caching and time allocation policy.

\begin{Def} [Feasible Joint Policy] Consider  a
joint caching and time allocation policy $(\mathbf c, \mathbf T_u, \mathbf T_d)$, where the caching design $\mathbf c$ does not change with the system state $(\mathbf X,\mathbf H)$, and the  time allocation design $(\mathbf T_u, \mathbf T_d)$ is a vector mapping (i.e., function) from the system state $(\mathbf X,\mathbf H)$ to the time allocation action $(\mathbf t_u, \mathbf t_d)$, i.e., $ \mathbf t_u=\mathbf T_u(\mathbf X,\mathbf H)$ and $ \mathbf t_d=\mathbf T_d(\mathbf X,\mathbf H)$.  Here,  $\mathbf T_u\triangleq (T_{u,n})_{n\in\mathcal X}$ and  $\mathbf T_d\triangleq (T_{d,n})_{n\in\mathcal X}$.  We call a policy  $(\mathbf c, \mathbf T_u, \mathbf T_d)$ feasible, if the caching design $\mathbf c$ satisfies \eqref{caching} and  \eqref{eqn:cache-const},  and the  time allocation action $(\mathbf t_u, \mathbf t_d)$ at each system state $(\mathbf X,\mathbf H)$ together with $\mathbf c$ satisfies \eqref{eqn:time_constraint_1_}, \eqref{eqn:time_constraint_2_}  and \eqref{eqn:time_constraint_3_}.
\label{Def:policy}
\end{Def}


\begin{remark} [Interpretation of Definition~\ref{Def:policy}]
Caching is  in general  in a much larger time-scale (e.g., on an hourly or daily   basis) and should reflect statistics of the system. In contrast, time allocation is in a much shorter time-scale (e.g., miliseconds) and should exploit instantaneous information of the system. Thus, in Definition~\ref{Def:policy}, we assume that the caching design depends only on the p.m.f.s $p(\mathbf x,\mathbf h)$, $(\mathbf x,\mathbf h)\in \mathcal X^K\times \mathcal H^K$ and  does not change with $(\mathbf X,\mathbf H)$, while the time allocation design is adaptive to $(\mathbf X,\mathbf H)$. In addition,  in this paper, we ignore the cost for placing the computation results into the storage at the BS in the initial stage, as the computation results may be useful for much longer time and the initial cost is negligible.
\end{remark}

Denote the  set of feasible joint policies    by $\Pi$.
Under a feasible joint policy  $(\mathbf c, \mathbf T_u, \mathbf T_d)\in \Pi$, the   average total energy  is given by
\begin{align}\label{eqn:expectation E}
\overline{E}(\mathbf c, \mathbf T_u, \mathbf T_d)\triangleq \mathbb E\left[ E(\mathbf c,\mathbf T_u(\mathbf X,\mathbf H),\mathbf T_d(\mathbf X, \mathbf H), \mathbf X,\mathbf H)\right],
\end{align}
where the expectation $\mathbb E$ is taken over the random system state $(\mathbf X, \mathbf H)\in \mathcal X^K\times \mathcal H^K $ and $E(\mathbf c,\mathbf T_u(\mathbf X,\mathbf H),\mathbf T_d(\mathbf X, \mathbf H), \mathbf X,\mathbf H)$ is given by \eqref{eqn:energy-X-H}. From \eqref{eqn:expectation E}, we can see that the joint   policy $(\mathbf c, \mathbf T_u, \mathbf T_d)$ significantly affects the average total energy.

In this paper, we would like to obtain the optimal joint feasible policy to minimize the average total energy. Specifically, we have the following optimization problem.

\begin{problem}[Average Total Energy Minimization]\label{prob:general-energy}
\begin{align}
\overline{E}^*\triangleq \min_{(\mathbf c, \mathbf T_u, \mathbf T_d)\in \Pi}&\quad \overline{E}(\mathbf c,\mathbf T_u,\mathbf T_d) \nonumber
\end{align}
Let $(\mathbf c^*,\mathbf T_u^*,\mathbf T_d^*)$  and $\overline{E}^*$ denote an optimal solution and  the optimal value, respectively.
\end{problem}

Problem~\ref{prob:general-energy} is a very challenging  mixed discrete-continuous optimization problem with two types of variables,  i.e.,  the caching design (discrete variables $\mathbf c$), and  the time allocation design (continuous variables $\mathbf T_u(\mathbf X, \mathbf H),(\mathbf X,\mathbf H)\in \mathcal X^K\times \mathcal H^K$ and $\mathbf T_d(\mathbf X, \mathbf H),(\mathbf X,\mathbf H)\in \mathcal X^K\times \mathcal H^K$). It can be shown  that Problem~\ref{prob:general-energy} is NP-hard.\footnote{For any given $(\mathbf T_u,\mathbf T_d)$, the minimization of $\overline{E}(\mathbf c,\mathbf T_u,\mathbf T_d) $ over $\mathbf c$  under the constraints in \eqref{caching} and  \eqref{eqn:cache-const} is a knapsack problem, which is NP-hard \cite{book04Hans}.}

Although  Problem~\ref{prob:general-energy} is for time interval $[0,T]$, the solution of Problem~\ref{prob:general-energy} can be applied to  a practical MEC system over a long time during which  the task popularity and channel statistics do not change. Specifically, the cached computation results can be used to satisfy task demands after time $T$. In addition, the time allocation design can be used for  a group of tasks that are generated at the same time after time $T$ and have the same deadline.

\begin{figure*}[!t]
\centering
{\resizebox{18cm}{!}{\includegraphics{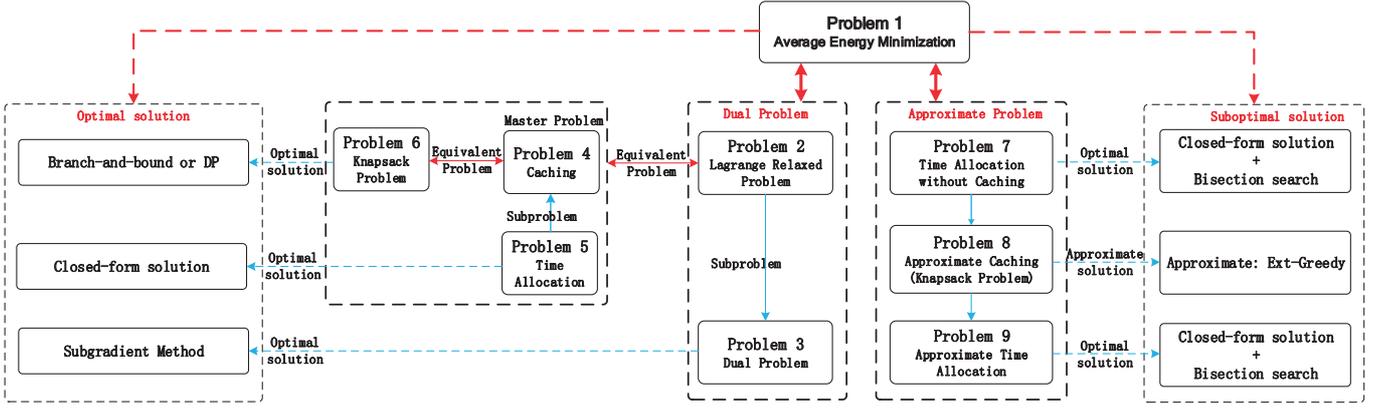}}}
\caption{\small{Proposed optimal and suboptimal solutions of Problem~\ref{prob:general-energy}.}}
\label{structure}\label{fig:flow}
\end{figure*}

\section{Optimal Solution}\label{Sec:opt}
In this section, we  obtain an optimal solution of  Problem~\ref{prob:general-energy} using  a dual method \cite{Bertsekasbooknonlinear:99}, as illustrated in Fig.~\ref{fig:flow}.


\subsection{Dual Problem}

One challenge in dealing with Problem~\ref{prob:general-energy} lies in the fact that   it is difficult to handle the deadline constraints for all $(\mathbf X,\mathbf H)\in \mathcal X^K\times \mathcal H^K$ in  \eqref{eqn:time_constraint_3_} (in terms of $T_{u,n}(\mathbf X,\mathbf H)$  and $T_{d,n}(\mathbf X,\mathbf H)$  instead of $t_{u,n}$ and $t_{d,n}$) where   $\mathbf c$ and  $(\mathbf T_u,\mathbf T_d)$ are coupled.
By eliminating the coupling constraints in  \eqref{eqn:time_constraint_3_} using nonnegative Lagrangian multipliers $\boldsymbol\lambda \triangleq (\lambda(\mathbf X, \mathbf H))_{(\mathbf X,\mathbf H)\in \mathcal X^K\times \mathcal H^K}\succeq 0$,\footnote{The notation $\succeq$ indicates the component-wise $\geq$.} we form the Lagrangian function $L(\mathbf c,\mathbf T_u,\mathbf T_d, \boldsymbol {\lambda})$ given in \eqref{eqn:Lagrangian}.
\begin{figure*}[!t]
\normalsize{
\begin{align}\label{eqn:Lagrangian}
&L(\mathbf c,\mathbf T_u,\mathbf T_d, \boldsymbol {\lambda})=\overline{E}(\mathbf c,\mathbf T_u,\mathbf T_d)+\sum_{(\mathbf X,\mathbf H)\in \mathcal X^K\times \mathcal H^K}\lambda(\mathbf X, \mathbf H)\left(\sum_{n \in \mathcal X}((1-c_n)T_{u,n}(\mathbf X, \mathbf H)+ T_{d,n}(\mathbf X, \mathbf H))- T\right)\end{align}
} \hrulefill
\end{figure*}
The dual function $g(\boldsymbol {\lambda})$ can be obtained by solving the following problem.
\begin{problem}[Lagrangian Relaxed Problem]\label{prob:Lagrangian Relaxed Problem} 
\begin{align}
g(\boldsymbol {\lambda})\triangleq\min_{\mathbf c,\mathbf T_u,\mathbf T_d}&\quad L(\mathbf c,\mathbf T_u,\mathbf T_d, \boldsymbol {\lambda})\nonumber\\
s.t.&\quad \eqref{caching},\eqref{eqn:cache-const},\eqref{eqn:time_constraint_1_}, \eqref{eqn:time_constraint_2_},\nonumber
\end{align}
where $L(\mathbf c,\mathbf T_u,\mathbf T_d, \boldsymbol {\lambda})$ is given by \eqref{eqn:Lagrangian}.  Let $(\tilde{\mathbf c}^*(\boldsymbol{\lambda}),\tilde{\mathbf T}^*_u(\boldsymbol{ \lambda}),\tilde{\mathbf T}^*_d(\boldsymbol{ \lambda}))$ denote an optimal solution.
\end{problem}

The dual problem   is given below.
\begin{problem}[Dual Problem]\label{prob:Lagrangian Dual Problem}
\begin{align}
g^*\triangleq \max_{\boldsymbol{\lambda}}&\quad g(\boldsymbol {\lambda})\nonumber\\
s.t.&\quad \boldsymbol{\lambda }\succeq 0,
\end{align}
where $g(\boldsymbol {\lambda})$ is given by Problem~\ref{prob:Lagrangian Relaxed Problem}.
Let $\boldsymbol{\lambda}^*$ and $g^*$ denote the optimal dual solution and the optimal dual value, respectively.
\end{problem}

By the weak duality theorem\cite{Bertsekasbooknonlinear:99},  $g^*\leq \overline{E}^*$,
where
$g^*$ is the optimal dual value of the dual problem in Problem~\ref{prob:Lagrangian Dual Problem},  and $\overline{E}^*$ is the optimal primal value of  the primal problem in Problem~\ref{prob:general-energy}. If $g^*=\overline{E}^*$ there is no duality gap (i.e., strong duality holds) and if $g^*<\overline{E}^*$ there is a duality gap.   The dual problem in Problem~\ref{prob:Lagrangian Dual Problem} is convex and   is more tractable than the primal problem in Problem~\ref{prob:general-energy}. Note that strong duality does not in general hold  for mixed discrete-continuous optimization problems.
If  we can obtain  the optimal dual solution $\boldsymbol{\lambda}^*$ and prove that there is no duality gap, an optimal primal solution $(\mathbf c^*,\mathbf T_u^*,\mathbf T_d^*)$ can be obtained by solving Problem~\ref{prob:Lagrangian Relaxed Problem} at $\boldsymbol{\lambda}^*$, i.e., $(\mathbf c^*,\mathbf T_u^*,\mathbf T_d^*)=(\tilde{\mathbf c}^*(\boldsymbol{\lambda}^*),\tilde{\mathbf T}^*_u(\boldsymbol{ \lambda}^*),\tilde{\mathbf T}^*_d(\boldsymbol{ \lambda}^*))$.

\subsection{Optimal Dual Solution}\label{subsec:dual-opt}
In this part, we  solve the dual problem in Problem~\ref{prob:Lagrangian Dual Problem}. First,  we need to obtain the dual function $g(\boldsymbol {\lambda})$ by solving Problem~\ref{prob:Lagrangian Relaxed Problem}. Note that Problem~\ref{prob:Lagrangian Relaxed Problem} is also a mixed discrete-continuous optimization problem with two types of variables,  i.e.,  the caching design (discrete variables $\mathbf c$), and  the time allocation (continuous variables $\mathbf T_u(\mathbf X, \mathbf H),(\mathbf X,\mathbf H)\in \mathcal X^K\times \mathcal H^K$ and $\mathbf T_d(\mathbf X, \mathbf H),(\mathbf X,\mathbf H)\in \mathcal X^K\times \mathcal H^K$). To facilitate the solution, we  equivalently  convert Problem~\ref{prob:Lagrangian Relaxed Problem} into a master problem and  multiple subproblems by separating the two types of variables and by noting that $L(\mathbf c,\mathbf T_u,\mathbf T_d, \boldsymbol {\lambda})$ and  $L_n(c,t_{u,n},t_{d,n},\mathbf X, \mathbf H, \lambda)$ satisfy \eqref{eqn:L-L-n}, where $L(\cdot)$ and  $L_n(\cdot)$ are  given by \eqref{eqn:Lagrangian} and \eqref{eqn:tilde-L-n}, respectively.
\begin{figure*}[!t]
\normalsize{
\begin{align}
&L(\mathbf c,\mathbf T_u,\mathbf T_d, \boldsymbol {\lambda})=\sum_{(\mathbf X,\mathbf H)\in \mathcal X^K\times \mathcal H^K}\sum_{n\in\mathcal X}L_{n}(c_n,  T_{u,n}(\mathbf X, \mathbf H), T_{d,n}(\mathbf X, \mathbf H), \mathbf X, \mathbf H, \lambda(\mathbf X, \mathbf H))-T\sum_{(\mathbf X,\mathbf H)\in \mathcal X^K\times \mathcal H^K}\lambda(\mathbf X,\mathbf H)\label{eqn:L-L-n}\\
&L_n(c,t_{u,n},t_{d,n},\mathbf X, \mathbf H,\lambda) \triangleq  p(\mathbf X, \mathbf H) E_n(c, t_{u,n},t_{d,n}, \mathbf X,\mathbf H)+\lambda((1-c)t_{u,n}+ t_{d,n})\label{eqn:tilde-L-n}
\end{align}
} \hrulefill
\end{figure*}
Specifically, the master problem is for the caching design and is given below.
\begin{problem}[Master Problem-Caching]\label{prob:caching decision} For all  $\boldsymbol \lambda \succeq 0$, we have
\begin{align*}
g(\boldsymbol \lambda )=\min_{\mathbf c}&\quad \sum_{\mathbf X \in \mathcal X^K}\sum_{\mathbf H \in \mathcal H^K} \sum_{n \in \mathcal X} L_{n}^*(c_n, \mathbf X, \mathbf H,\lambda(\mathbf X, \mathbf H))\\
s.t.&\quad~\eqref{caching},\eqref{eqn:cache-const},
\end{align*}
where $L_{n}^*(\cdot)$ is given by the following subproblem. Let $\tilde{\mathbf c}^*(\boldsymbol \lambda)\triangleq (\tilde{c}_n^*(\boldsymbol \lambda))_{n\in\mathcal X}$ denote the optimal solution.
\end{problem}

Each subproblem  is for the uploading and downloading time  allocation for one task at one system state, and is given below.
\begin{problem}[Subproblem-Time Allocation]\label{prob:time allocation} For all $(\mathbf X,\mathbf H)\in \mathcal X^K\times \mathcal H^K$, $\lambda$, $n\in \mathcal X$ and  $c\in\{0,1\}$,  we have
\begin{align*}
L_{n}^*(c,\mathbf X,\mathbf H,\lambda)\triangleq\min_{ t_{u,n},t_{d,n}}&\quad L_n(c,t_{u,n},t_{d,n},\mathbf X, \mathbf H, \lambda)\\
s.t.&\quad \eqref{eqn:time_constraint_1_},\eqref{eqn:time_constraint_2_},
\end{align*}
where $L_n(\cdot)$ is given by \eqref{eqn:tilde-L-n}.
Let $\tilde T_{u,n}^*(c,\mathbf X,\mathbf H,\lambda) $ and $\tilde T_{d,n}^*(c, \mathbf X,\mathbf H,\lambda)$ denote the optimal solution.
\end{problem}

First, we solve Problem~\ref{prob:time allocation}.  Problem~\ref{prob:time allocation} is convex and strong duality holds. Using KKT conditions, we can obtain the optimal solution of Problem~\ref{prob:time allocation}, which is given below.

\begin{lemma}[Optimal Solution of Problem~\ref{prob:time allocation}]\label{lemma:solve_t}
For all $(\mathbf X,\mathbf H)\in \mathcal X^K\times \mathcal H^K$, $\lambda$, $n\in \mathcal X$ and  $c\in\{0,1\}$, the optimal solution of Problem~\ref{prob:time allocation} is given by
\begin{align}
&\tilde{T}_{u,n}^*( c,\mathbf X,\mathbf H,\lambda)=(1-c) f(\mathbf X, \mathbf H,  L_{u,n},H_{u,n},\lambda),\label{eqn:tilde-t-u-n}\\
&\tilde{T}_{d,n}^*( c,\mathbf X,\mathbf H,\lambda)=f(\mathbf X, \mathbf H,L_{d,n},H_{d,n},\lambda),\label{eqn:tilde-t-d-n}
\end{align}
where $f(\cdot)$ is given by \eqref{eqn:f} with  $W(\cdot)$ being the Lambert function.
\begin{figure*}[!t]
\normalsize{\begin{align}
f(\mathbf X, \mathbf H,  x,y,\lambda) \triangleq& \min\left\{K_n(\mathbf X),1\right\}
\max\left\{\min\left\{
\frac{x \ln2}{B\left(W\left(\frac{\frac{\lambda y}{p(\mathbf X, \mathbf H)} - n_0}{n_0 e}\right) + 1 \right)} ,T\right\},0\right\}\label{eqn:f}\\
e_{1,n}(\mathbf X,\mathbf H,\lambda)=&p(\mathbf X, \mathbf H)(E_{u,n}(f(\mathbf X, \mathbf H,  L_{u,n},H_{u,n},\lambda),\mathbf X,\mathbf H) +E_{e,n}(\mathbf X))+\lambda f(\mathbf X, \mathbf H,  L_{u,n},H_{u,n},\lambda)\label{eqn:e1n}\\
e_{2,n}(\mathbf X,\mathbf H,\lambda)=&p(\mathbf X, \mathbf H)E_{d,n}(f(\mathbf X, \mathbf H,L_{d,n},H_{d,n},\lambda),\mathbf X,\mathbf H)+\lambda f(\mathbf X, \mathbf H,L_{d,n},H_{d,n},\lambda)\label{eqn:e2n}\
\end{align}
} \hrulefill
\end{figure*}
Furthermore, the optimal value of Problem~\ref{prob:time allocation} is $$L_n^*(c,\mathbf X,\mathbf H,\lambda)=(1-c)e_{1,n}( \mathbf X, \mathbf H,\lambda)+e_{2,n}( \mathbf X, \mathbf H,\lambda),$$
where $e_{1,n}(\cdot)$ and $e_{2,n}(\cdot)$ are given by \eqref{eqn:e1n} and \eqref{eqn:e2n}, respectively.
\end{lemma}

Next, we solve Problem~\ref{prob:caching decision}. We introduce the following knapsack problem.
\begin{problem} [Knapsack Problem for Caching]For all  $\boldsymbol \lambda \succeq 0$, we have  \begin{align*}
\max_{\mathbf c}&\quad\sum_{n \in \mathcal X}c_n\sum_{\mathbf X \in \mathcal X^K}\sum_{ \mathbf H \in \mathcal H^K}e_{1,n}(\mathbf X,\mathbf H,\lambda(\mathbf X, \mathbf H))\\
s.t.&\quad~\eqref{caching},\eqref{eqn:cache-const},
\end{align*}
where $e_{1,n}(\cdot)$ is  given by \eqref{eqn:e1n}.\label{prob:knapsack}
\end{problem}

By exploring structural properties of Problem~\ref{prob:caching decision}, we have have the following result.
\begin{lemma}[Equivalence between Problem~\ref{prob:caching decision} and  Problem~\ref{prob:knapsack}]\label{lemma cache}
An optimal solution of  Problem~\ref{prob:knapsack} is also optimal for Problem~\ref{prob:caching decision}.
\end{lemma}

By Lemma~\ref{lemma cache}, we can obtain  $\tilde{\mathbf c}^*(\boldsymbol \lambda)$ by solving  the knapsack problem in Problem~\ref{prob:knapsack}  instead of  Problem~\ref{prob:caching decision}. Note that knapsack problem is an NP-hard problem and can be solved optimally using two approaches, i.e., the branch-and-bound method and dynamic programming (DP), with non-polynomial complexity \cite{book04Hans}.
Substituting    $\tilde{\mathbf c}^*(\boldsymbol{\lambda})$  into the optimal solution of Problem~\ref{prob:time allocation} in \eqref{eqn:tilde-t-u-n} and \eqref{eqn:tilde-t-d-n}, we have $\left(\tilde{T}_{u,n}^*(\tilde c_n^*(\boldsymbol{\lambda}),\mathbf X,\mathbf H,\lambda(\mathbf X,\mathbf H)), \tilde{T}_{d,n}^*(\tilde c_n^*(\boldsymbol{\lambda}),\mathbf X,\mathbf H,\lambda(\mathbf X,\mathbf H))\right)$. With abuse of notation, denote with $(\tilde{T}_{u,n}^*(\boldsymbol{\lambda}),\tilde{T}_{d,n}^*(\boldsymbol{\lambda}))$ the corresponding mapping. Let $\tilde{\mathbf T}_u^*(\boldsymbol{\lambda})\triangleq (\tilde{T}_{u,n}^*(\boldsymbol{\lambda}))_{n\in\mathcal X}$ and $\tilde{\mathbf T}_d^*(\boldsymbol{\lambda})\triangleq (\tilde{T}_{d,n}^*(\boldsymbol{\lambda}))_{n\in\mathcal X}$. Thus, we can obtain an optimal solution of Problem~\ref{prob:Lagrangian Relaxed Problem}, i.e., $(\tilde{\mathbf c}^*(\boldsymbol{\lambda}), \tilde{\mathbf T}_u^*(\boldsymbol{\lambda}), \tilde{\mathbf T}_d^*(\boldsymbol{\lambda}))$.
Furthermore, we can obtain the optimal value of Problem~\ref{prob:Lagrangian Relaxed Problem}, i.e., the dual function $g(\boldsymbol {\lambda})=L(\tilde{\mathbf c}^*(\boldsymbol{\lambda}),\tilde{\mathbf T}^*_u(\boldsymbol{ \lambda}),\tilde{\mathbf T}^*_d(\boldsymbol{ \lambda}), \boldsymbol {\lambda})$.

Finally, we solve the dual problem in Problem~\ref{prob:Lagrangian Dual Problem}. As there typically exist some Lagrangian multipliers  for which Problem~\ref{prob:Lagrangian Dual Problem} has multiple optimal solutions, the dual function $g(\boldsymbol {\lambda})$   is non-differentiable, and gradient methods cannot be applied to solve Problem~\ref{prob:Lagrangian Dual Problem}. Here, we consider the subgradient method which uses subgradients as directions of improvement of the distance to the optimum \cite{Bertsekasbooknonlinear:99}.
In particular, for all $(\mathbf X,\mathbf H)\in \mathcal X^K\times \mathcal H^K$, the subgradient  method generates a sequence of dual feasible points according to the following iteration:
\begin{align}
&\lambda_{t+1}(\mathbf X, \mathbf H)=\max\left\{\lambda_t(\mathbf X, \mathbf H)+\alpha_ts(\mathbf X, \mathbf H,\boldsymbol\lambda_t),0\right\},\label{eqn:update}
\end{align}
where $s(\mathbf X, \mathbf H,\boldsymbol\lambda_t)$ denotes  a subgradient of $g(\boldsymbol {\lambda}_{t})$  given by:
\begin{align}
&s(\mathbf X, \mathbf H,\boldsymbol\lambda_t)\nonumber\\=&\sum_{n \in \mathcal X}\Big((1-\tilde c_n^*(\boldsymbol{\lambda}_t))\tilde T^*_{u,n}(\tilde c_n^*(\boldsymbol{\lambda}_t),\mathbf X, \mathbf H,  \lambda_t(\mathbf X, \mathbf H)) \nonumber\\&+T^*_{d,n}(\tilde c_n^*(\boldsymbol{\lambda}_t),\mathbf X, \mathbf H,  \lambda_t(\mathbf X, \mathbf H))\Big)- T.\label{eqn:s}
\end{align}
Here, $t$ is the iteration index and $\alpha_t$ is the
step-size, e.g., $\alpha_t=(1+m)/(t+m)$, where $m$ is a fixed nonnegative number. Note that the updates of $\lambda_{t+1}(\mathbf X, \mathbf H), (\mathbf X,\mathbf H)\in \mathcal X^K\times \mathcal H^K$ are coupled through $\tilde{\mathbf c}^*(\boldsymbol{\lambda})$. It has been shown  in  \cite{Bertsekasbooknonlinear:99} that  $\boldsymbol \lambda_t\to \boldsymbol{\lambda}^*$  as $t \to \infty$ for all initial points $\boldsymbol{\lambda}_0 \succeq 0$.
Therefore, using the subgradient method, we can obtain the dual optimal solution $\boldsymbol{\lambda}^*$.

\subsection{Optimal Primal Solution} \label{subsec:primal-opt}

Problem \ref{prob:general-energy} is a mixed discrete-continuous optimization problem, for which  strong duality does not in general  hold. By analyzing structural properties, we show that strong duality holds for  Problem \ref{prob:general-energy}.
\begin{theorem} [Strong Duality] $g^*=\overline{E}^*$ holds and $(\mathbf c^*,\mathbf T_u^*,\mathbf T_d^*)=(\tilde{\mathbf c}^*(\boldsymbol{\lambda}^*), \tilde{\mathbf T}_u^*(\boldsymbol{\lambda}^*), \tilde{\mathbf T}_d^*(\boldsymbol{\lambda}^*))$.\label{Thm:strongduality}
\end{theorem}

Theorem~\ref{Thm:strongduality} indicates that   the primal optimal solution $(\mathbf c^*,\mathbf T_u^*,\mathbf T_d^*)$ can be obtained by the above-mentioned dual method.

In summary, we can obtain an optimal solution $(\mathbf c^*,\mathbf T_u^*,\mathbf T_d^*)$  by repeating three steps, i.e., solving the caching design problem in  Problem~\ref{prob:knapsack} (which relies on the optimal solution of the time allocation problem in  Problem~\ref{prob:time allocation}) for given $\boldsymbol \lambda_t$, solving the time allocation problem in Problem~\ref{prob:time allocation} based on  the obtained caching design, and updating $\boldsymbol \lambda_t$ based on  the obtained caching design and the time allocation design, until $\boldsymbol \lambda_t$ converges or stopping criterion is satisfied. The details for obtaining the optimal solution are summarized in Algorithm~\ref{alg:opt}.

\begin{algorithm}
    \caption{\small{Optimal Algorithm}}
\begin{small}
     \begin{algorithmic}[1]
           \STATE Set iteration index $t=0$, and initialize $\boldsymbol{\lambda}_t$.
           \REPEAT
           \STATE Obtain $\tilde{\mathbf c}^*(\boldsymbol{\lambda}_t)$ by solving Problem~\ref{prob:knapsack} using branch-and-bound method or DP.
           \STATE For all $(\mathbf X,\mathbf H)\in \mathcal X^K\times \mathcal H^K$ and $n\in\mathcal X$, compute  $\tilde{T}_{u,n}^*(\tilde c_n^*(\boldsymbol{\lambda}),\mathbf X,\mathbf H,\lambda(\mathbf X,\mathbf H))$ and $\tilde{T}_{d,n}^*(\tilde c_n^*(\boldsymbol{\lambda}),\mathbf X,\mathbf H,\lambda(\mathbf X,\mathbf H))$ according to \eqref{eqn:tilde-t-u-n} and \eqref{eqn:tilde-t-d-n}, respectively.
           \STATE For all $(\mathbf X,\mathbf H)\in \mathcal X^K\times \mathcal H^K$, compute $\lambda_{t+1}(\mathbf X, \mathbf H)$ according to \eqref{eqn:update}, where $s(\mathbf X, \mathbf H,\boldsymbol\lambda_t)$ is obtained according to \eqref{eqn:s}.
           \STATE Set $t=t+1$.
           \UNTIL{stopping criterion  (e.g., $|s(\mathbf X, \mathbf H,\boldsymbol\lambda_{t-1})|<\epsilon$, where $\epsilon$ is small and positive) is satisfied.}
    \end{algorithmic}
    \end{small}\label{alg:opt}
\end{algorithm}


\section{Low-Complexity Suboptimal Solution}

From \eqref{eqn:update}, we see that  $\lambda_{t+1}(\mathbf X, \mathbf H)$, $(\mathbf X, \mathbf H)\in\mathcal X^K\times \mathcal H^K$ all depend on $\boldsymbol \lambda_{t}$ via $\tilde{\mathbf c}^*(\boldsymbol{\lambda}_t)$. That is, the updates of $\lambda_{t}(\mathbf X, \mathbf H)$, $(\mathbf X, \mathbf H)\in\mathcal X^K\times \mathcal H^K$ are coupled. Thus,  $\boldsymbol \lambda_t$ may converge  to $\boldsymbol{\lambda}^*$ slowly,  leading to high computational complexity for obtaining an optimal solution $(\mathbf c^*,\mathbf T_u^*,\mathbf T_d^*)$ using the dual method in Section~\ref{Sec:opt}, especially when the system state space $\mathcal X^K\times \mathcal H^K$ is large. In this section, as illustrated in Fig.~\ref{fig:flow}, we obtain a low-complexity suboptimal solution by carefully handling the coupling among all $(\mathbf X,\mathbf H)\in \mathcal X^K\times \mathcal H^K$ which results from the coupling between the caching design and the time allocation design. Specifically, instead of joint optimization, we optimize the two designs separately.

Before obtaining  a suboptimal caching design,  we first ignore storage resource (i.e., by setting $C=0$ and $\mathbf c=\mathbf 0$) and consider  Problem~\ref{prob:general-energy} with $C=0$  (i.e., minimizing $ \overline{E}(\mathbf 0,\mathbf T_u,\mathbf T_d) $ over all feasible $(\mathbf T_u, \mathbf T_d)$ with $(\mathbf 0, \mathbf T_u, \mathbf T_d)\in \Pi$). This problem can be  equivalently separated into  the following  time allocation problems without caching, one for each $(\mathbf X,\mathbf H)\in \mathcal X^K\times \mathcal H^K$.

\begin{problem}[Time Allocation without Caching]\label{prob:time allocation-nocache} For all $(\mathbf X,\mathbf H)\in \mathcal X^K\times \mathcal H^K$,  we have
\begin{align}
\min_{ \mathbf t_{u},\mathbf t_{d}}&\quad \sum_{n\in\mathcal X}E_n(0, t_{u,n},t_{d,n}, \mathbf X,\mathbf H)\nonumber\\
s.t.&\quad \eqref{eqn:time_constraint_1_},\eqref{eqn:time_constraint_2_}, \nonumber\\
&\quad \sum_{n\in\mathcal X}\left(t_{u,n} + t_{d,n} \right)\leq T,\nonumber
\end{align}
where $E_n(\cdot)$ is given by \eqref{eqn:energy-X-H-n}.
Let $\left( \mathbf T_u^{0\dagger}(\mathbf X, \mathbf H),  \mathbf T_d^{0\dagger}(\mathbf X, \mathbf H)\right)$ denote the optimal solution, where $\mathbf T_u^{0\dagger}\triangleq (T_{u,n}^{0\dagger})_{n\in\mathcal X}$ and $\mathbf T_d^{0\dagger}\triangleq (T_{d,n}^{0\dagger})_{n\in\mathcal X}$.
\end{problem}

Problem~\ref{prob:time allocation-nocache} is convex and strong duality holds. Similarly, using KKT conditions, we can obtain the optimal solution of Problem~\ref{prob:time allocation}:
\begin{align}
&T_{u,n}^{0\dagger}(\mathbf X,\mathbf H)=f(\mathbf X, \mathbf H,  L_{u,n},H_{u,n},\lambda^{0\dagger}(\mathbf X, \mathbf H)),\label{eqn:t-u-n-dagger}\\
&T_{d,n}^{0\dagger}(\mathbf X,\mathbf H)=f(\mathbf X, \mathbf H,L_{d,n},H_{d,n},\lambda^{0\dagger}(\mathbf X, \mathbf H)),\label{eqn:t-d-n-dagger}
\end{align}
where $f(\cdot)$ is given by \eqref{eqn:f} with  $W(\cdot)$ being the Lambert function and $\lambda^{0\dagger}(\mathbf X, \mathbf H)$ satisfies $$\sum_{n\in\mathcal X}\left(T_{u,n}^{0\dagger}(\mathbf X,\mathbf H) + T_{d,n}^{0\dagger}(\mathbf X,\mathbf H)\right)= T.$$
As  $f(\cdot )$ in \eqref{eqn:f} is a non-increasing function of $\lambda$, $\lambda^{0\dagger}(\mathbf X, \mathbf H)$ can be easily obtained using bisection search.

Then, we take the storage resource into consideration and focus on caching only, i.e,  obtaining an optimal caching design which minimizes $ \overline{E}(\mathbf c,\mathbf T_u^{0\dagger},\mathbf T_d^{0\dagger}) $ subject to \eqref{caching} and \eqref{eqn:cache-const}. Similarly, this is equivalent to consider the following knapsack problem, which is NP-hard.
\begin{problem} [Approximate Knapsack Problem for Caching]\begin{align*}
\max_{\mathbf c}&\quad\sum_{n \in \mathcal X}c_n\sum_{\mathbf X \in \mathcal X^K}\sum_{ \mathbf H \in \mathcal H^K}e_{1,n}(\mathbf X,\mathbf H,\lambda^{0\dagger}(\mathbf X, \mathbf H))\\
s.t.&\quad~\eqref{caching},\eqref{eqn:cache-const},
\end{align*}
where $e_{1,n}(\cdot)$ is  given by \eqref{eqn:e1n}.
\label{prob:approx-knapsack}
\end{problem}

An approximate solution with $1/2$ optimality guarantee and polynomial complexity  can be obtained using the Ext-Greedy algorithm proposed in \cite{book04Hans}.
Based on the suboptimal solution  denoted by $\mathbf c^{\dagger}\triangleq (c^{\dagger}_n)_{n\in\mathcal X}$, we then focus on the optimal time allocation design which minimizes $ \overline{E}(\mathbf c^{\dagger},\mathbf T_u,\mathbf T_d) $ over all feasible $(\mathbf T_u, \mathbf T_d)$ with $(\mathbf c^{\dagger}, \mathbf T_u, \mathbf T_d)\in \Pi$. Similarly, this problem can be  equivalently separated into  the following  time allocation problems for  the given caching design $\mathbf c^{\dagger}$, one for each $(\mathbf X,\mathbf H)\in \mathcal X^K\times \mathcal H^K$.

\begin{problem}[Approximate Time Allocation]\label{prob:appx-time allocation} Given $\mathbf c^{\dagger}$, for all $(\mathbf X,\mathbf H)\in \mathcal X^K\times \mathcal H^K$,  we have
\begin{align}
\min_{ \mathbf t_{u},\mathbf t_{d}}&\quad \sum_{n\in\mathcal X}E_n(c_n^{\dagger}, t_{u,n},t_{d,n}, \mathbf X,\mathbf H)\nonumber\\
s.t.&\quad \eqref{eqn:time_constraint_1_},\eqref{eqn:time_constraint_2_}, \nonumber\\
&\quad \sum_{n\in\mathcal X}\left((1-c_n^{\dagger})t_{u,n} + t_{d,n} \right)\leq T,
\nonumber
\end{align}
where $E_n(\cdot)$ is given by \eqref{eqn:energy-X-H-n}.
Let $\left( \mathbf T_u^{\dagger}(\mathbf X, \mathbf H),  \mathbf T_d^{\dagger}(\mathbf X, \mathbf H)\right)$ denote the optimal solution, where $\mathbf T_u^{\dagger}\triangleq (T_{u,n}^{\dagger})_{n\in\mathcal X}$ and $\mathbf T_d^{\dagger}\triangleq (T_{d,n}^{\dagger})_{n\in\mathcal X}$.
\end{problem}

Similarly, we can obtain the optimal solution of Problem~\ref{prob:appx-time allocation}:
\begin{align}
&T_{u,n}^{\dagger}(\mathbf X,\mathbf H)=(1-c_n^{\dagger})f(\mathbf X, \mathbf H,  L_{u,n},H_{u,n},\lambda^{\dagger}(\mathbf X, \mathbf H)),\label{eqn:t-u-n-dagger-cache}\\
&T_{d,n}^{\dagger}(\mathbf X,\mathbf H)=f(\mathbf X, \mathbf H,L_{d,n},H_{d,n},\lambda^{\dagger}(\mathbf X, \mathbf H)),\label{eqn:t-d-n-dagger-cache}
\end{align}
where $f(\cdot)$ is given by \eqref{eqn:f} with  $W(\cdot)$ being the Lambert function and $\lambda^{\dagger}(\mathbf X, \mathbf H)$ satisfies  $$\sum_{n\in\mathcal X}\left((1-c_n^{\dagger})T_{u,n}^{\dagger}(\mathbf X,\mathbf H) + T_{d,n}^{\dagger}(\mathbf X,\mathbf H)\right)=T.$$
$\lambda^{\dagger}(\mathbf X, \mathbf H)$ can be easily obtained using bisection search.

In summary, we can obtain a suboptimal solution $(\mathbf c^{\dagger},\mathbf T_u^{\dagger},\mathbf T_d^{\dagger}) $ by sequentially solving the approximate caching design problem in Problem~\ref{prob:approx-knapsack} (which relies on the optimal solution of the time allocation problem without caching in Problem~\ref{prob:time allocation-nocache}) and  the approximate time allocation problem in Problem~\ref{prob:appx-time allocation}. In obtaining the suboptimal solution,   for any $(\mathbf X,\mathbf H)\in \mathcal X^K\times \mathcal H^K$, both $\lambda^{0\dagger}(\mathbf X, \mathbf H)$ and $\lambda^{\dagger}(\mathbf X, \mathbf H)$ are obtained using efficient bisection search, there is no coupling among $(\mathbf X,\mathbf H)\in \mathcal X^K\times \mathcal H^K$, and no iterations are required in this process.
The details for obtaining the suboptimal solution are summarized in Algorithm~\ref{alg:subopt}. It is clear that Algorithm~\ref{alg:subopt} has much lower computational complexity than Algorithm~\ref{alg:opt}.

\begin{algorithm}
    \caption{\small{Low-complexity Suboptimal Algorithm}}
\begin{small}
     \begin{algorithmic}[1]
     \STATE For all $(\mathbf X,\mathbf H)\in \mathcal X^K\times \mathcal H^K$, compute $\lambda^{0\dagger}(\mathbf X, \mathbf H)$ by solving $\sum_{n\in\mathcal X}\left(T_{u,n}^{0\dagger}(\mathbf X,\mathbf H) + T_{d,n}^{0\dagger}(\mathbf X,\mathbf H)\right)= T$ via bisection search;
     \STATE Compute $\mathbf c^{\dagger}$ by solving Problem~\ref{prob:approx-knapsack} using the Ext-Greedy algorithm;
     \STATE For all $(\mathbf X,\mathbf H)\in \mathcal X^K\times \mathcal H^K$, compute $\lambda^{\dagger}(\mathbf X, \mathbf H)$ by solving $\sum_{n\in\mathcal X}\left((1-c_n^{\dagger})T_{u,n}^{\dagger}(\mathbf X,\mathbf H) + T_{d,n}^{\dagger}(\mathbf X,\mathbf H)\right)=T$ via bisection search, and  for all $n\in\mathcal X$, compute $T_{u,n}^{\dagger}(\mathbf X,\mathbf H)$ and $T_{d,n}^{\dagger}(\mathbf X,\mathbf H)$ according to \eqref{eqn:t-u-n-dagger-cache} and \eqref{eqn:t-d-n-dagger-cache}, respectively.
    \end{algorithmic}
    \end{small}\label{alg:subopt}
\end{algorithm}

\begin{figure}[t]
\begin{center}
\subfigure{\resizebox{6.8cm}{!}{\includegraphics{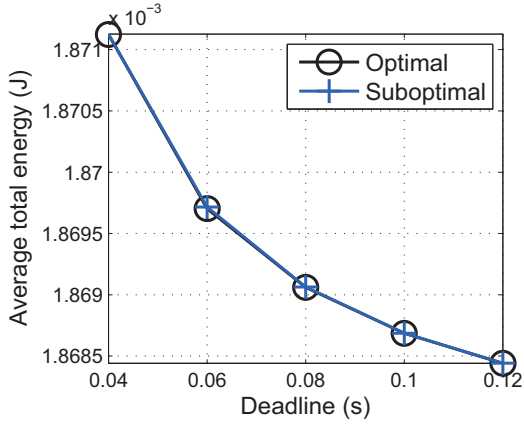}}}
\end{center}
\vspace*{-0.3cm}
   \caption{\small{Average total energy versus deadline $T$ at $\gamma=0.8$, $C=5\times10^4$ bits, $K=2, N=3$.} }
   \label{fig:opt-suba}
   \vspace*{-0.5cm}
\end{figure}
\begin{figure}[t]
\begin{center}
\subfigure{\resizebox{6.35cm}{!}{\includegraphics{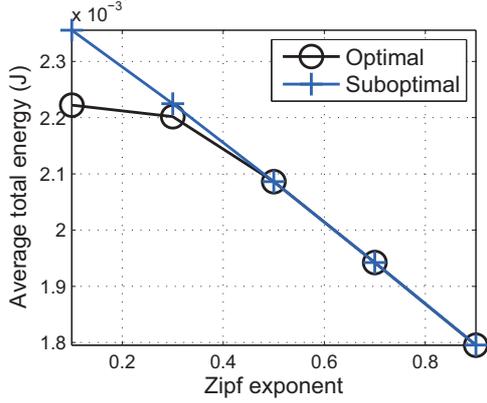}}}
\end{center}
\vspace*{-0.3cm}
   \caption{\small{Average total energy versus Zipf exponent $\gamma$ at $T=0.08$ s, $C=5\times10^4$ bits, $K=2, N=3$.} }
   \label{fig:opt-subb}
   \vspace*{-0.2cm}
\end{figure}

\section{Numerical Results}
In the numerical experiment, we consider the following settings\cite{you2016energy}. Let $B =10$ MHz, $n_0 = 10^{-9}$ W, $\mu=10^{-30}$, $F_b=6\times10^9$, $L_{u,n}=n\times4\times10^4+1\times10^4$ bits, $L_{d,n}=n\times2\times10^4+1\times10^4$ bits and $L_{e,n}=n\times4\times10^4+1\times10^4$ CPU-cycles, for all  $n\in\mathcal{X}$.
{Set $\mathcal{H}=\{5\times10^{-7},1.5\times10^{-6}\}$ and $p_{H_k}(5\times10^{-7})=0.7015$,  $p_{H_k}(1.5\times10^{-6})=0.2581$ for all $k\in\mathcal{K}$.}
Assume that $X_k, k\in\mathcal K$ follow  the same Zipf distribution, i.e.,
$p_{X_k}(n)=\frac{n^{-\gamma}}{\sum_{i\in\mathcal{X}}i^{-\gamma}}$ for all $k\in\mathcal K$, where $\gamma$ is the Zipf exponent.
\subsection{Comparison Between Optimal and Suboptimal Solutions}

In this part, we compare the proposed optimal and  suboptimal solutions at small $K$ and $N$ so that the computational complexity for obtaining the optimal solution is manageable.  From Fig.~\ref{fig:opt-suba} and Fig.~\ref{fig:opt-subb}, we can see that the average total energy  of the proposed suboptimal solution is very close to that of the optimal solution, demonstrating its applicability at small $K$ and $N$.

\subsection{Comparisons with Existing Schemes}

In this part, we compare the proposed suboptimal solution with four baseline schemes\cite{you2016energy}.  All the four baseline schemes view the tasks from different mobiles as different tasks and consider the uploading and downloading of these tasks separately. In addition, Baseline~1 and Baseline~2 make use of the storage resource and adopt the same caching design as the proposed suboptimal solution, while Baseline~3 and Baseline~4 do not consider the caching of computation results.   Baseline~1 and Baseline~3 consider equal uploading and downloading time allocation among $K$ mobiles, i.e.,
\begin{align}
    &t_{u,n}=t_{d,n}=
    \begin{cases}
    \frac{T}{\sum_{i\in\{j\in\mathcal{X}:K_{j}(\mathbf{X})\geq 1\}}(2-c_i)}, & K_n(\mathbf X)\geq 1\\
    0, &\text{otherwise}
\end{cases}\nonumber
\end{align}
where $c_n=c_n^{\dagger}$ for Baseline~1 and $c_n=0$ for Baseline~3, for all $n\in \mathcal X$.
Baseline~2 and Baseline~4 allocate the  uploading and downloading time durations for the task of each mobile proportionally  to the sizes of its task input  and computation result, respectively, i.e.,
\begin{align}
    &t_{u,n}=
    \begin{cases}
   \frac{L_{u,n}T}{\sum_{i\in\{j\in\mathcal{X}:K_{j}(\mathbf{X})\geq 1\}}((1-c_i)L_{u,i}+L_{d,i})}, & K_n(\mathbf X)\geq 1\\
    0, &\text{otherwise}
\end{cases}\nonumber\\
    &t_{d,n}=
    \begin{cases}
   \frac{L_{d,n}T}{\sum_{i\in\{j\in\mathcal{X}:K_{j}(\mathbf{X})\geq 1\}}((1-c_i)L_{u,i}+L_{d,i})}, & K_n(\mathbf X)\geq 1\\
    0, &\text{otherwise}
\end{cases}\nonumber
\end{align}
where $c_n=c_n^{\dagger}$ for Baseline~2 and $c_n=0$ for Baseline~4, for all $n\in \mathcal X$.

\begin{figure}[t]
\begin{center}
\subfigure{\resizebox{6.6cm}{!}{\includegraphics{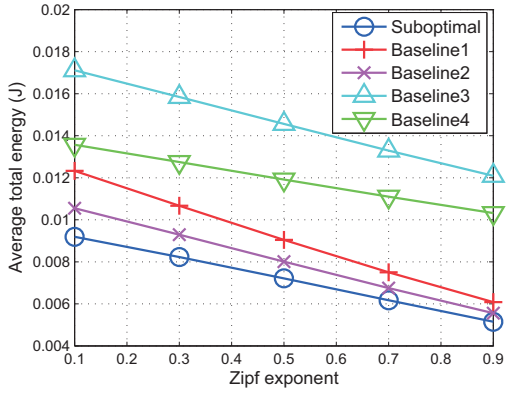}}}\
\end{center}
\vspace*{-0.3cm}
   \caption{\small{Average total energy versus Zipf exponent $\gamma$ at $C=2.4\times10^5~\text{bits}, K=4, N=12$, $T=0.08$~\text{s}.}}
   \label{fig:schemesa}
   \vspace*{-0.2cm}
\end{figure}

\begin{figure}[t]
\begin{center}
  \subfigure{\resizebox{6.6cm}{!}{\includegraphics{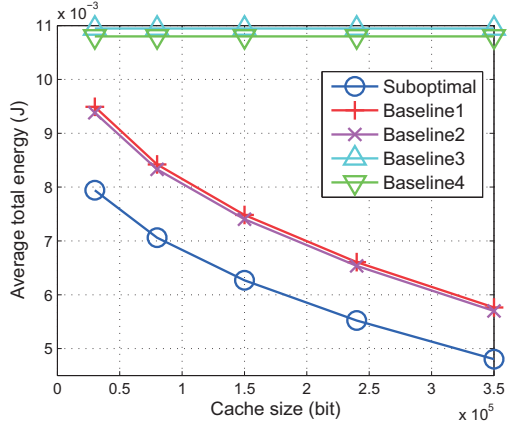}}}
\end{center}
\vspace*{-0.3cm}
   \caption{\small{Average total energy versus cache size $C$ at $\gamma=0.8, K=4, N=12$, $T=0.08$~\text{s}.}}
   \label{fig:schemesb}
   \vspace*{-0.2cm}
\end{figure}

\begin{figure}[t]
\begin{center}
  \subfigure{\resizebox{6.6cm}{!}{\includegraphics{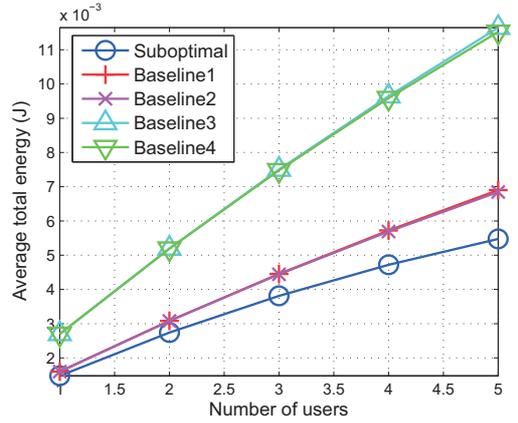}}}\
\end{center}
\vspace*{-0.3cm}
   \caption{\small{Average total energy versus number of users $K$ at  $\gamma=0.8, C=1.5\times10^5~\text{bits}, N=9$, $T=0.08$~\text{s}.}}
   \label{fig:schemesc}
   \vspace*{-0.2cm}
\end{figure}

\begin{figure}[t]
\begin{center}
   \subfigure{\resizebox{6.6cm}{!}{\includegraphics{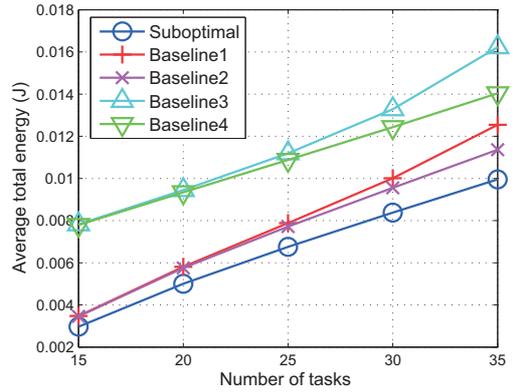}}}
\end{center}
\vspace*{-0.3cm}
   \caption{\small{Average total energy versus number of tasks $N$ at $\gamma=0.8, C=3.5\times10^5~\text{bits}, K=3$, $T=0.08$~\text{s}.}}
   \label{fig:schemesd}
   \vspace*{-0.2cm}
\end{figure}

 Fig.~\ref{fig:schemesa},  Fig.~\ref{fig:schemesb},  Fig.~\ref{fig:schemesc} and  Fig.~\ref{fig:schemesd} illustrate the energy consumption versus different  parameters.
From Fig.~\ref{fig:schemesa},  Fig.~\ref{fig:schemesb},  Fig.~\ref{fig:schemesc} and  Fig.~\ref{fig:schemesd}, we can observe that the proposed suboptimal solution outperforms the four baselines, demonstrating the advantage of the proposed suboptimal solution in efficiently  utilizing the storage and communication resources.
When $\gamma$ increases, 
the average total energy of each scheme decreases, as the effective task load reduces.
When $C$ increases, the average total energies of the proposed suboptimal solution,  Baseline~1 and Baseline~2 decrease, due to the energy reduction in task   executing.
When $K$ or  $N$ increases,  the average total energy of each scheme increases, due to the increase of the computation load.
The performance gains of the proposed suboptimal solution over Baseline~1 and Baseline~2 come from the the fact that the proposed suboptimal solution  avoids redundant transmissions in uploading and downloading. Baseline~1 and Baseline~2 outperform Baseline~3 and Baseline~4, respectively, by making use of the storage resource.

\section{Conclusion}
In this paper, we  consider the average total energy minimization problem
subject to the caching and deadline constraints
to optimally allocate the
storage resource at the BS for caching computation results as
well as the uploading and downloading time durations  in a multi-user cache-assisted MEC system.
The problem is a challenging mixed discrete-continuous optimization problem.
We show that strong duality holds, and obtain an optimal solution
using a dual method. We further propose a low-complexity suboptimal solution.
Finally, numerical results show that the proposed suboptimal
solution outperforms existing comparison schemes and reveal the advantage in efficiently utilizing storage and communication resources.
This paper provides key insights for  designing  energy-efficient MEC systems by jointly utilizing communication, caching and computation.

\end{document}